\documentclass[a4,11pt]{cip-submit}  
\usepackage{afterpage}

\usepackage{mathptmx}
\usepackage{hyperref}
\hypersetup{
	colorlinks=false,
	pdfborder={0 0 0}
}
\def\Mr{\uppercase}

\usepackage{float}
\usepackage{wasysym}
\usepackage{graphicx,wrapfig,tikz}
\usepackage{caption}
\usepackage{subcaption}
\usepackage{amsmath,amsxtra,amssymb,latexsym,amscd,color,bm}
\usepackage[mathscr]{eucal}
\usepackage{array}
\usepackage{cite}
\def\nA{nucleon-nucleus\ }

\def\vsm{\vskip0.1cm}

\def\titles#1{\title{\large\bf\noindent #1}}
\def\authors#1{\author{\begin{flushleft}{#1}\end{flushleft}}}
\def\authord#1#2{\indent\Mr{#1}\\
	\textit{\indent#2}\vsm}

\def\email#1{\bigskip\href{mailto:#1}{\textit{E-mail:}~{#1}}\\[3mm]}

\def\Keywords#1{\\[.2cm] Keywords:~{#1}.}

\begin{document}
\Year{2018} \Page{1}\Endpage{10}
\titles{$R$-MATRIX METHOD AND THE NONLOCAL NUCLEON OPTICAL POTENTIAL}
\author{
\authord{DOAN THI LOAN, NGUYEN HOANG PHUC, DAO TIEN KHOA}
 {Institute for Nuclear Science and Technology, VINATOM \\
 179 Hoang Quoc Viet, Cau Giay, Hanoi, Vietnam}}
\maketitle
\markboth{D.~T. LOAN, N.~H. PHUC, D.~T. KHOA}{$R$-MATRIX METHOD ANG THE NONLOCAL 
 NUCLEON OPTICAL POTENTIAL}
\begin{abstract}
The calculable R-matrix method is applied to solve the Schr\"odinger equation 
in the optical model (OM) analysis of the elastic \nA scattering using a nonlocal 
nucleon optical potential (OP). The phenomenological nonlocal nucleon OP proposed 
by Perey and Buck (PB), and the two recent versions of the PB parametrization 
were used in the present OM study of the elastic nucleon scattering on $^{27}$Al, 
$^{40}$Ca, $^{48}$Ca, $^{90}$Zr, and $^{208}$Pb targets at different energies. 
The comparison of the OM results given by the calculable $R$-matrix method with 
those given by other methods confirms that the calculable $R$-matrix method is 
an efficient tool for the OM study of the elastic \nA scattering using a nonlocal 
nucleon OP.
\end{abstract}
\section{\Mr{Introduction}}
\label{intro}
The nucleon-nucleus scattering remains an important experiment of the modern
nuclear physics to investigate the \nA interaction as well as 
the structure of the target nucleus. In particular, the elastic and inelastic 
scattering of the short-lived, unstable nuclei on proton target, or proton scattering 
in the inverse kinematics, is now extensively carried out with the secondary beams 
of unstable nuclei to investigate the unknown structure of unstable nuclei under 
study. The key quantity needed for the description of the \nA scattering 
at both low and high energies is the nucleon optical potential (OP), which 
determines the scattering wave function of the scattered nucleon as solution 
of the Schr\"{o}dinger equation in a single-channel optical model (OM) or 
of the system of the coupled-channel (CC) equations. The elastic scattering wave 
function given by the OP is often dubbed as the distorted wave in the Distorted 
Wave Born Approximation (DWBA) or CC formalism, which are widely used to study 
different processes of the direct nuclear reaction \cite{Sat1}. 
Therefore, a proper treatment of the nucleon OP is always of a vital importance. 

Over the years, for the simplicity of the numerical calculation, the OP was mainly 
assumed in the local form for the OM analysis of the elastic \nA scattering. 
The phenomenological Woods-Saxon (WS) form is mostly used in numerous 
parametrizations of the local nucleon OP \cite{CH89, KD}. However, it is well known 
that the nucleon OP is nonlocal in the coordinate space due to the Pauli 
principle and multichannel coupling. In a microscopic theory of the nucleon 
OP, one can obtain the formal expression for the complex nucleon OP 
using the projection technique developed by Feshbach \cite{Fe92}
\begin{eqnarray}
U = V_0+\lim_{\epsilon \rightarrow 0}\sum_{\alpha\alpha'}V_{0\to\alpha}
\left(\frac{1}{E-H+i\epsilon}\right)_{\alpha\alpha'}V_{\alpha'\to 0}, \label{Feshbach}
\end{eqnarray}
where the first-order term $V_0$ is the interaction between the incident nucleon 
and the target nucleus being in its ground state (g.s.), while $V_{0\to\alpha}$ is 
that when the target nucleus is excited to a state labeled by $\alpha$ which, 
in general, can be also in the continuum or a breakup channel. The first term of 
Eq.~(\ref{Feshbach}) is real and can be obtained by the folding model 
\cite{Ken01,Br77,Kho02} as a Hartree-Fock-type potential using the realistic 
wave function for the target g.s. and a proper effective nucleon-nucleon 
interaction. If the antisymmetrization of the \nA system is taken exactly into 
account \cite{Ken01}, then the Fock term of $V_0$ is nonlocal in the coordinate 
space. The second term of Eq.~(\ref{Feshbach}) is often referred to as the dynamic 
polarization potential (DPP), which arises from the coupling between the elastic 
scattering channel and the (open) nonelastic channels. The DPP is complex and 
nonlocal because the target nucleus that is excited at position $\bm{r}$ can return 
to its g.s. at another position $\bm{r}'$, with $\bm{r}\neq\bm{r'}$ \cite{Bran}. 
Thus, the nonlocality of the nucleon OP has a justified physics origin and it is 
of importance to implement the use of the nonlocal OP in the OM, DWBA, and CC studies 
of the \nA scattering.

Although most of the OM studies of the elastic \nA scattering have used the local
nucleon OP, several studies have been carried out using the nucleon OP in
an explicit nonlocal form (see, e.g., Refs.~\cite{Perey, NLOM, Kosho, TPM, Lovel}).  
Among them, we note the early work by Perey and Buck (PB) \cite{Perey} 
and the recent revision of the PB parametrization by Tian, Pang, and Ma (TPM) 
\cite{TPM}, which used a nonlocal nucleon OP built up from a WS form 
factor multiplied by a nonlocal Gaussian. While the PB parameters were adjusted 
to the best OM description of the two data sets (elastic $n+^{208}$Pb scattering 
at 7.0 and 14.5 MeV), those of the TPM potential were fitted to reproduce the 
data of the elastic nucleon scattering on $^{32}$S, $^{56}$Fe, $^{120}$Sn, and 
$^{208}$Pb targets at energies of 8 to 30 MeV. More recently, an energy dependence 
of the nonlocal OP was introduced explicitly into the imaginary parts of the PB 
and TPM potentials, dubbed as PB-E and TPM-E potentials, whose parameters were 
fitted to the best OM description of the nucleon elastic scattering data on $^{40}$Ca, 
$^{90}$Zr and $^{208}$Pb targets at energies $E\approx 5 - 45$ MeV \cite{Lovel,Lovel2}.

In general, solving the Schr\"{o}dinger equation with a nonlocal potential readily
leads to an integro-differential equation, which is more complicated than a standard
differential equation with a local potential. For the \nA scattering problem, 
the use of the nonlocal OP leads to an explicit angular-momentum dependence 
of the integral equation for the scattering wave function. At variance 
with the traditional methods usually used for the integro-differential equation, 
we apply the calculable $R$-matrix method \cite{desco} in the present work to solve 
exactly the Schr\"{o}dinger equation with a nonlocal nucleon OP. The recent extension 
of the calculable $R$-matrix method \cite{desco, desco2} has included the Lagrange 
mesh and Gauss-Legendre quadrature integration that simplify the numerical calculation 
significantly. Although the $R$-matrix method was developed to treat exactly
the nonlocal (central) potential, it has been applied mainly to study the
nuclear resonant scattering at low energies using some local form for the 
scattering potential \cite{desco}. The aim of the present work is, therefore, 
to explore the applicability of the calculable $R$-matrix method in the OM 
calculation of the elastic \nA scattering using a nonlocal nucleon OP.   
For the nonlocal nucleon OP, we have chosen the PB \cite{Perey} and 
TPM \cite{TPM} potentials as well as the energy-dependent nonlocal OP by Lovell 
{\it et al.} \cite{Lovel,Lovel2} to study the elastic nucleon scattering on $^{27}$Al, 
$^{40}$Ca, $^{48}$Ca, $^{90}$Zr, and $^{208}$Pb targets at energies of 14.6 MeV 
to 40 MeV. The OM results obtained with the calculable $R$-matrix method are 
compared with those obtained with other methods given by the computer codes NLOM
\cite{NLOM,NLOM2}, NLAT \cite{Titus2}, and DWBA98 \cite{Raynal}. 

\section{Elastic nucleon scattering and the $R$-matrix method}
\subsection{Optical model with a nonlocal nucleon OP}
We give here a brief introduction into the OM calculation with a nucleon OP that 
contains a nonlocal central term \cite{Perey,NLOM,TPM,Lovel}. In this case, 
the scattering wave function $\chi$ of the incident nucleon is obtained by solving 
the following Schr\"odinger equation 
\begin{eqnarray}
\left[-\frac{\hbar^2}{2\mu}\nabla^2+V_{\rm C}(\bm{r})+V_{\rm so}(\bm{r})
\bm{l}.{\bm{\sigma}}\right]\chi(\bm{k},\bm{r})+\!\!
\int V(\bm{r},\bm{r}')\chi(\bm{k},\bm{r}')d\bm{r}'=E\chi(\bm{k},\bm{r}).  \label{eq1}
\end{eqnarray}
Here $V_{\rm C}(\bm{r})$ and $V_{\rm so}(\bm{r})$ are the Coulomb and spin-orbit 
potentials, respectively, which are assumed to be local, and $V(\bm{r},\bm{r}')$ 
is the nonlocal central potential. The Coulomb potential for the incident 
protons is given as
\begin{eqnarray}
V_{\rm C}(r)=\left\{\begin{array}{lcr}
\displaystyle\frac{Ze^2}{r},& & r > R_{\rm C}\\
\displaystyle\frac{Ze^2}{2R_{\rm C}}\left(3-\frac{r^2}{R_{\rm C}^2}\right),& & 
 r\leqslant R_{\rm C}.
\end{array}\right. 
\end{eqnarray}
The Coulomb radius $R_{\rm C}$ is determined by the mass-number dependent
formula, and the spin-orbit potential $V_{\rm so}(\bm{r})$ is adopted in the usual 
Thomas form \cite{CH89}. The wave number is determined by the reduced mass $\mu$ and 
center-of-mass energy $E$ as $k=\sqrt{2\mu E/\hbar^2}$. The spin of the incident 
nucleon is $\bm{s}=\hbar\bm{\sigma}/2$, where $\bm{\sigma}$ is the Pauli matrix. 
In the OM calculation of the elastic nucleon scattering, the orientation of the 
nucleon spin needs to be treated explicitly, and the scattering wave function 
(also dubbed as the distorted wave) is expressed \cite{Sat1} via the nucleon 
spinor $\xi$ as 
\begin{equation}
\chi_{m_s}(\bm{k},\bm{r})=\sum_{m'_s}\chi_{m_sm'_s}(\bm{k},\bm{r})\xi_{\frac{1}{2}m'_s},
\end{equation}
where the spin matrix elements of the distorted wave are given by the following
partial-wave series 
\begin{equation}\label{2}
\chi_{m_sm'_s}(\bm{k},\bm{r})=\frac{4\pi}{kr}\sum_{ljm_j}\psi_{lj}(k,r)\langle 
l\frac{1}{2}mm_s|jm_j\rangle\langle l\frac{1}{2}m'm'_s|jm_j\rangle
[i^lY_{lm'}(\hat{\bm{r}})]Y^*_{lm}(\hat{\bm{k}}).
\end{equation}
Here $Y_{lm}$ are the spherical harmonics with $m=m_j-m_s$ and $m'=m_j-m'_s$.
The nonlocal central potential can also be expanded over the spherical harmonics 
series as
\begin{equation}
 V(\bm{r},\bm{r'})=\sum_{lm}\frac{v_l(r,r')}{rr'}[i^lY_{lm}(\theta,\phi)]
 Y^*_{lm}(\theta',\phi').
\end{equation}
Then, the radial equation for $\psi_{lj}(k,r)$ is readily obtained by multiplying 
both sides of Eq.~(\ref{eq1}) with the spherical harmonics $Y_{lm}(\hat{\bm{k}})$ 
and $Y_{lm'}(\hat{\bm{r}})$, and integrating out the angular variables 
$\hat{\bm{r}}$ and $\hat{\bm{k}}$.
\begin{eqnarray} \label{pareq}
-\frac{\hbar^2}{2\mu}\left[\frac{d^2}{dr^2}-\frac{l(l+1)}{r^2}\right]\psi_{lj}(k,r)+ 
 \left[V_{\rm C}(r)+ A_{lj}V_{\rm so}(r)\right]\psi_{lj}(k,r)\nonumber\\
 +\int v_l(r,r')\psi_{lj}(k,r')dr'=E\psi_{lj}(k,r),
\end{eqnarray}
where $A_{lj}=l$ when $j=l+\frac{1}{2}$, and $A_{lj}=-l-1$ when $j=l-\frac{1}{2}$. 
The spin matrix elements $\chi_{m_sm'_s}(\bm{r},\bm{k})$ of the distorted wave  
obtained from the solutions $\psi_{lj}(k,r)$ of Eq.~({\ref{pareq}}) can be expressed 
in terms of the Coulomb $\chi_{\rm C}$ and scattered $\chi_{\rm scatt}$ waves as
\begin{equation}
\chi_{m_sm'_s}(\bm{k},\bm{r})=\chi_{\rm C}(\bm{k},\bm{r})\delta_{m_sm'_s}
+\chi_{\text{scatt},m_sm'_s}(\bm{k},\bm{r}),
\end{equation}
and $\chi_{\text{scatt},m_sm'_s}$ has the following asymptotic form 
\begin{equation}
\chi_{\text{scatt},m_sm'_s}(\bm{k},\bm{r})\sim 
\frac{e^{i(kr-\eta\ln 2kr)}}{r}f_{m_sm'_s}(\theta),
\end{equation}
where $f_{m_sm'_s}(\theta)$ are the spin matrix elements of the elastic scattering 
amplitude, $\eta$ is the usual Coulomb parameter. Matching $\psi_{lj}$ with the 
corresponding term of the Coulomb wave function at large radii, where the nuclear 
potential is negligible, we obtain the scattering phase shift $\delta_l^\pm$. Then,
the partial-wave elements of the scattering matrix ($S_{lj}=S_l^+$ with 
$j=l+\frac{1}{2}$, and $S_{lj}=S_l^-$ with $j=l-\frac{1}{2}$) are 
determined by the relation
\begin{equation}
 S_l^{\pm}=\exp({2i\delta_l^{\pm}}).
\end{equation}
The amplitude of the elastic nucleon-nucleus scattering can be expressed 
in terms of an operator $\bm{f}$ acting on the spin of the incident nucleon 
\cite{Sat1} as
\begin{equation}
\bm{f}(\theta)= g(\theta)\bm{1}+ih(\theta)\bm{\sigma}.\bm{n},
\end{equation}
where $\bm{n}$ is the unit vector along the direction $\bm{k}\times\bm{k'}$, which is
perpendicular to the scattering plane. We further adopt the frame with the $z$-axis 
aligned along $\bm{k}$, and the $y$-axis along $\bm{n}$. Then the spin matrix elements 
of $\bm{f}$ can be expressed as
\begin{equation}
 f_{++}=f_{--}=g(\theta), f_{+-}=-f_{-+}=h(\theta),
\end{equation}
where the subscript $\pm$ denotes $m_s=\pm\frac{1}{2}$. The explicit expressions 
for $g(\theta)$ and $h(\theta)$ functions are \cite{Sat1}  
\begin{eqnarray}
&&g(\theta)=f_{\rm C}(\theta)+\frac{i}{2k}\sum_l[(2l+1)-(l+1)S^+_l-lS^-_l]
 \exp(2i\sigma_l)P^0_l(\cos\theta)\nonumber\\
&&h(\theta)=\frac{i}{2k}\sum_l(S^-_l-S^+_l)\exp(2i\sigma_l)P^1_l(\cos\theta),
\end{eqnarray}
where $f_{\rm C}(\theta)$ and $\sigma_l$ are the Rutherford scattering amplitude 
and Coulomb phase shift, respectively, and $P^{0(1)}_l(\cos\theta)$ are 
the Legendre polynomials. Finally, the differential scattering cross section for the 
elastic (unpolarized) nucleon scattering is obtained as
\begin{eqnarray}
\frac{d\sigma(\theta)}{d\Omega}=\frac{1}{2}\sum_{m_sm'_s}|f_{m_sm'_s}(\theta)|^2=
|g(\theta)|^2+|h(\theta)|^2.\label{cros}
\end{eqnarray}

\subsection{The calculable $R$-matrix method}
The calculable $R$-matrix method introduced in Ref.~\cite{desco} is an efficient tool 
to solve the scattering problem using the Sch\"{o}dinger equation. It is different 
from the phenomenological $R$-matrix method which is a technique to parametrize 
the cross section of the nuclear resonant scattering. The main principle of solving 
the Schr\"odinger Eq.~(\ref{pareq}) using the calculable $R$-matrix method is the 
division of the configuration space at the channel radius $a$ into an internal region 
and an external region. The channel radius $a$ is chosen large enough so that 
the nuclear potential is negligible in the external region. In the present work, 
$a = 15 fm$ was chosen for all the OM calculations using the $R$-matrix method. 
The partial-wave component $\psi_{lj}(r)$ of the scattering wave function in the 
external region can then be written as
\begin{equation}\label{14}
\psi_{lj}^{\rm ext}(r)=\frac{i}{2}\left[I_l(kr)-S_{lj}O_l(kr)\right],
\end{equation}
with the conjugate functions $I_l$ and $O_l$ determined as
\begin{equation}
I_l=G_l-iF_l,\  O_l=G_l+iF_l,
\end{equation}
where $F_l$ and $G_l$ are the regular and irregular Coulomb functions, respectively.
In the internal region, the wave function is expanded over some finite basis of 
$N$ linearly independent function $\varphi_n$ as
\begin{equation}\label{15}
\psi_{lj}^{\rm int}(r)=\sum_{n=1}^{N}c_n\varphi_n(r).
\end{equation}
The internal and external parts of the radial wave function can be connected 
at the boundary $r=a$ through the continuity condition $\psi^{\rm{int}}_{lj}(a)
=\psi^{\rm{ext}}_{lj}(a)=\psi_{lj}(a)$. This leads to the definition 
of the $R$-matrix at a given energy $E$ as
\begin{equation}\label{16}
\psi_{lj}(a)=\mathscr{R}_{lj}(E)[a\psi^{'}_{lj}(a)-B\psi_{lj}(a)],
\end{equation}
where $B$ is the dimensionless boundary parameter introduced for the later convenience. 
The $R$-matrix has dimension 1 in the single-channel case and is 
just a function of energy $E$. Because the Hamiltonian is not Hermitian over 
the internal region $(0, a)$, the following surface Bloch operator \cite{Bloch} 
is introduced
\begin{equation}\label{17}
\mathscr{L}(B)=\frac{\hbar^2}{2\mu}\delta(r-a)\left(\frac{d}{dr}-\frac{B}{r}\right),
\end{equation}
so that the combination of the Hamiltonian and Bloch operator is Hermitian over the 
region $(0, a)$ when $B$ is real. It is well known that for the scattering problem,
the results obtained using the $R$-matrix method are independent of $B$ \cite{desco}. 
Therefore, we have fixed $B = 0$ in the present work, and the Schr\"odinger equation 
in the internal region can be approximated by an inhomogeneous Bloch-Schr\"odinger 
equation. Eq.~(\ref{pareq}) can then be rewritten as
\begin{eqnarray}
\left\{-\frac{\hbar^2}{2\mu}\left[\frac{d^2}{dr^2}+
\frac{l(	l+1)}{r^2}\right]+V_{\rm C}(r)+A_{lj}
V_{\rm so}(r)-E+\mathscr{L}\right\}\sum_{i=1}^Nc_n\varphi_n(r)\nonumber\\
+\int v_l(r,r')\sum_{i=1}^Nc_n\varphi_n(r')dr'=\mathscr{L}\psi_{lj}^{\rm ext}(r).
\label{18}
\end{eqnarray}
The Bloch operator $\mathscr{L}$ ensures the continuity of the derivative of the wave 
function. Projecting both sides of Eq.~(\ref{18}) on $\varphi_i(r)$ and integrating 
over $r$ variable, we obtain
\begin{equation}\label{19}
\sum_{n=1}^NC_{in}(E)c_n=\frac{\hbar^2}{2\mu}\varphi_i(a)
\frac{d\psi_{lj}^{\rm ext}(r)}{dr}\Big|_{\displaystyle{r=a}},
\end{equation}
where the matrix elements $C_{in}(E)$ are determined as
\begin{eqnarray}\label{20}
C_{in}(E)\!\!\!&=&\!\!\!\!\!\int\varphi_i(r)\Big\{-\frac{\hbar^2}{2\mu}\left[\frac{d^2}{dr^2}+
\frac{l(l+1)}{r^2}\right]+V_{\rm C}(r) + A_{lj}V_{\rm so}(r) \nonumber\\
&& \!\!\!\!\! -E+\mathscr{L}\Big\}\varphi_n(r)dr 
+ \int \varphi_i(r)v_l(r,r')\varphi_n(r')drdr'.
\end{eqnarray}
The coefficients $c_n$ are obtained by solving the system of linear equations (\ref{19}). 
Inserting them into Eq.~(\ref{15}) at $r=a$ and using the boundary condition (\ref{16}),
we obtain the calculable $R$-matrix 
\begin{equation}
\mathscr{R}_{lj}(E)=\frac{\hbar^2}{2\mu a}\sum_{i,n=1}^N 
\varphi_i(a)(C^{-1})_{in}\varphi_n(a).
\end{equation}
For the convenience in the calculation of $C_{in}(E)$, the modified Lagrange 
functions are chosen as the basis functions $\varphi_n(r)$ (see the explicit 
expressions in Ref.~\cite{desco}). Using the Gauss-Legendre quadrature, the integral 
of any regular function can be approximated by a sum of the function values 
at the mesh points $\{r_i\}$ given by the solutions of the Legendre polynomial, 
multiplied by the weight $\lambda_i$ in the interval $[0,a]$. Then, the basic 
functions satisfy the Lagrange condition \cite{desco},
\begin{equation}
\varphi_n(r_i)=(\lambda_i)^{-1/2}\delta_{in}\label{Lag}.
\end{equation} 
Thanks to the relation (\ref{Lag}), the calculation is simplified significantly 
and we obtain for the all local potentials in Eq.~({\ref{20})
\begin{equation}\label{26}
\int_0^a \varphi_i(r) V(r)\varphi_n(r)dr= 
\sum_{k=1}^N \lambda_k \varphi_i(r_k)V(r_k)\varphi_n(r_k)= V(r_i)\delta_{in},
\end{equation}
and for the nonlocal central potential 
\begin{equation}\label{27}
\int \varphi_i(r) v_l(r,r') \varphi_n(r') dr'dr = v_l(r_i,r_n).
\end{equation}
Thus, the calculable $R$-matrix method allows us to obtain the solution of the 
scattering equation (\ref{18}) with a nonlocal central potential without using 
an iterative procedure. 
It is obvious that external wave functions contain the partial-wave 
elements $S_{lj}$ of the scattering matrix and internal wave functions which 
include the $R$-matrix component $\mathscr{R}_{lj}(E)$. Through the continuity 
condition of the scattering wave function, the relationship between $S_{lj}$ and 
$\mathscr{R}_{lj}(E)$ is obtained as 
\begin{equation}\label{29}
S_{lj}=\exp(2i\phi_l)\frac{1-L_l^*\mathscr{R}_{lj}(E)}{1-L_l}
\mathscr{R}_{lj}(E),
\end{equation}
where 
\begin{equation}
L_l=\frac{ka}{O_l(ka)}\frac{dO_l(kr)}{dr}\Big|_{\displaystyle{r=a}} \nonumber
\end{equation}
is the dimensionless logarithmic derivative of $O_l$ at the channel radius $a$, and 
\begin{equation}\label{31}
\phi_l=\arg I_l(ka)=-\arctan\left[\frac{F_l(ka)}{G_l(ka)}\right]
\end{equation}
is the hard-sphere Coulomb phase shift. 
The differential scattering cross section (\ref{cros}) for the elastic (unpolarized) 
nucleon scattering is then readily obtained from the partial-wave elements 
of the scattering matrix (\ref{29})
\subsection{The nonlocal nucleon OP}
To explore the use of the calculable $R$-matrix method, we adopt in the present work
the phenomenological nonlocal nucleon OP proposed by Perey and Buck \cite{Perey} 
for the neutron elastic scattering from $^{208}$Pb target at the energies of 7.0 and 
14.5 MeV. The original PB parameters have been improved recently by Tian, Pang, and 
Ma \cite{TPM} to reproduce the elastic nucleon scattering data on $^{32}$S, 
$^{56}$Fe, $^{120}$Sn, and $^{208}$Pb targets at the energies of 8 to 30 MeV. 
Like the original PB potential that consists of a nonlocal real volume term, 
a nonlocal imaginary surface term, and a local real spin-orbit potential, the TPM 
potential includes further a nonlocal imaginary volume term.  
More recently, the energy dependence of the nonlocal OP has been introduced 
explicitly into the imaginary parts of the PB and TPM potentials (the PB-E and 
TPM-E potentials) based on the OM fit to the neutron elastic scattering data 
on $^{40}$Ca, $^{90}$Zr, and $^{208}$Pb targets at the energies of 5 to 40 MeV 
\cite{Lovel}. All these versions of the nonlocal nucleon OP were parametrized in
the following form
\begin{equation}\label{8}
V(\bm{r},\bm{r'})=U\left(\frac{|\bm{r}+\bm{r'}|}{2}\right)
H\left(\frac{|\bm{r}-\bm{r'}|}{\beta}\right),
\end{equation}
where $\beta$ is the range of the nonlocality. The function 
$H(|\bm{r}-\bm{r'}|/\beta)$ was chosen \cite{Perey} in the Gaussian form 
\begin{equation}
 H\left(\frac{|\bm{r}-\bm{r'}|}{\beta}\right)=\frac{1}{\pi^\frac{3}{2}\beta^3}
 \exp\left[-\left(\frac{\bm{r}-\bm{r'}}{\beta}\right)^2\right],
\end{equation}
and the function $U(p)$ with $p=|\bm{r}+\bm{r'}|/2$ was chosen in the 
Woods-Saxon form like those usually used \cite{CH89} for the local OP 
\begin{equation}
-U(p) = V_{\rm R}f_{\rm R}(p)+iW_{\rm I} f_{\rm I}(p)+iW_{\rm D} f_{\rm D}(p),
\end{equation}
where
\begin{eqnarray}
f_{\rm R(I)}(p)&=&\left[1+\exp\left(\frac{p-R_{\rm R(I)}}
 {a_{\rm R(I)}}\right)\right]^{-1}  \\ \nonumber
f_{\rm D}(p)&=&4\exp\left(\frac{p-R_{\rm D}}{a_{\rm D}}\right)
 \left[1+\exp\left(\frac{p-R_{\rm D}}{a_{\rm D}}\right)\right]^{-2}.
\end{eqnarray}
$R_i=r_iA^{1/3}$ and $a_i$ ($i={\rm R,I,D}$) are the potential radius and diffuseness 
tabulated in Refs.~\cite{Perey,TPM,Lovel}. Because of the dominant contribution 
of the Gaussian factor to the integral at $\bm{r}\approx \bm{r'}$, the approximation
$p=|\bm{r}+\bm{r'}|/2\approx (r+r')/2$ is made \cite{Perey} to simplify the computation 
of $v_l(r,r')$. Then, one obtains
\begin{equation}\label{11}
v_l(r,r')=U\left(\frac{r+r'}{2}\right)\frac{1}{\pi^{\frac{1}{2}}\beta}
 \exp\left[-\frac{(r^2+r'^2)}{\beta^2}\right]2i^lzj_l(-iz),
\end{equation}
where $z=2rr'/\beta^2$ and $j_l(x)$ is the spherical Bessel function. The Schr\"odinger
equation (\ref{pareq}) using the nonlocal potential (\ref{18}) for the elastic 
nucleon scattering can be solved numerically by the traditional methods 
or by the calculable $R$-matrix method discussed above. 
\section{\Mr{Results and discussion}}
Before presenting the OM results obtained with the calculable $R$-matrix method,
it is of interest to discuss briefly the iterative method used in the OM
calculations with the nonlocal nucleon OP. In these calculations, a trial
local potential $U_{\text{init}}$ is used  to generate the initial scattering 
wave function for the iteration scheme \cite{Perey}
\begin{equation}
\frac{\hbar^2}{2\mu}\left[\frac{d^2}{dr^2}-\frac{l(l+1)}{r^2}\right]
\psi^{(0)}_{lj}(r)+\left[E-V_{\rm C}(r)- A_{lj}
V_{\rm so}(r)-U_{\text{init}}(r)\right] \psi^{(0)}_{lj}(r) = 0.
\end{equation}
The solution $\psi^{(0)}_{lj}(r)$ given by the initial local potential is then 
used in the first step ($n=1$) of solving the integro-differential 
equation iteratively    
\begin{eqnarray}
\frac{\hbar^2}{2\mu}\left[\frac{d^2}{dr^2}-\frac{l(l+1)}{r^2}\right]
\psi^{(n)}_{lj}(r)+\left[E-V_{\rm C}(r)- A_{lj}
V_{\rm so}(r)-U_{\text{init}}(r)\right] \psi^{(n)}_{lj}(r) \nonumber\\
= \int v_l(r,r')\psi^{(n-1)}_{lj}(r')dr' - U_{\text{init}}(r)\psi^{(n-1)}_{lj}(r).
\label{Iteq}
\end{eqnarray}
Such an iterative procedure performs as many iterations as necessary for the 
convergence of the scattering wave function. Although the choice of $U_{\text{init}}(r)$ 
does not affect the final solution of the scattering equation, the convergence
of the iterative method somewhat depends on the choice of the initial local potential. 
In the study of the nucleon transfer reactions by Titus {\it et al.} \cite{Titus2,Titus}, 
the iterative method has been used to determine both the scattering and bound nucleon wave 
functions (the code NLAT). 
\begin{figure}[bht]
\vspace{-1.0cm}
\hspace{-1.0cm}
\includegraphics[scale=0.41]{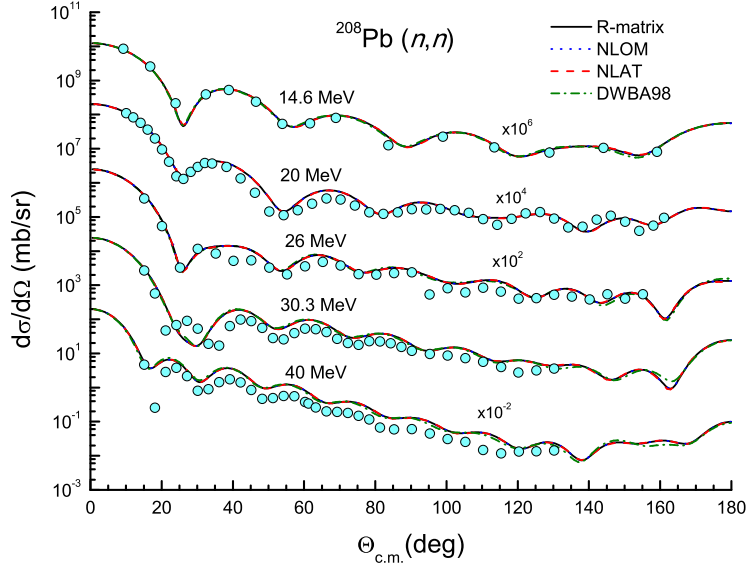} \vspace{-0.4cm}
\caption{OM descriptions of the neutron elastic scattering on $^{208}$Pb 
target at 14.6, 20, 26, 30.3 and 40 MeV given by the calculable $R$-matrix method 
and other methods (see detailed discussion in text) using the nonlocal 
PB potential \cite{Perey}. The experimental data were taken from 
Refs.~\cite{exp1,exp2,exp3,exp4,exp5}.} \label{fig1}
\vspace{-0.1cm}
\end{figure}

The NLOM method \cite{NLOM, NLOM2} was suggested to solve the nonlocal OM equation
quickly and reliably, using the Green-function transformation to convert the
scattering wave function to the form of a bound-state wave function that can be
expanded over an orthonormal basis. Then the main problem is to solve a set of the
inhomogeneous linear equations using, e.g., the Lanczos method which also needs an 
iterative procedure. In this method, the trial local potential is still necessary, 
but the convergence is not sensitive to the chosen trial potential.
\begin{figure}[bht]
\vspace{-1.0cm}
\hspace{-1.0cm} 
\includegraphics[scale=0.41]{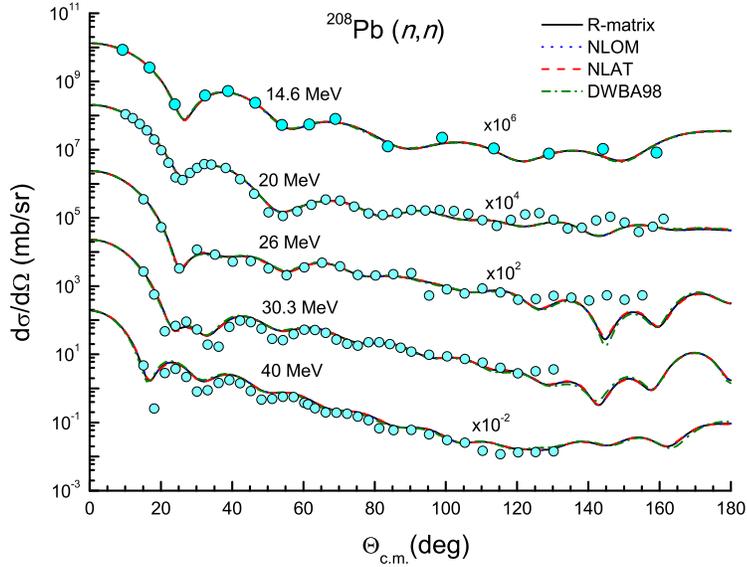} \vspace{-0.4cm}
\caption{The same at Fig.~\ref{fig1} but using the nonlocal TPM potential 
\cite{TPM}. The experimental data were taken from Refs.~\cite{exp1,exp2,exp3,exp4,exp5}.} 
\label{fig2}
\end{figure}
\begin{figure}[bht]
\vspace{-1.0cm}
\hspace{-1.0cm}
\includegraphics[scale=0.41]{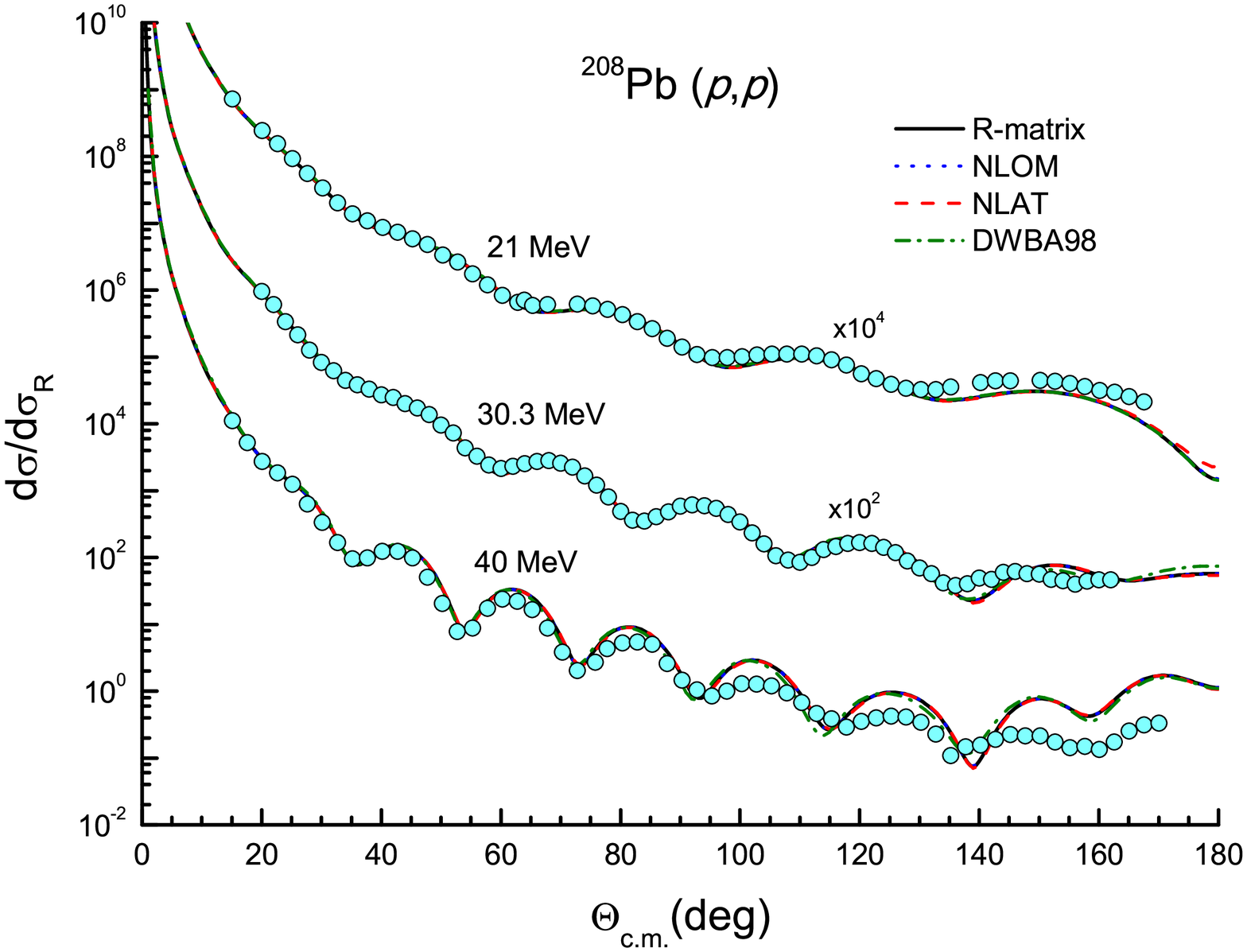} \vspace{-0.5cm}
\caption{OM descriptions of the proton elastic scattering on $^{208}$Pb 
target at 21, 30.3, and 40 MeV given by the calculable $R$-matrix method 
and other methods (see detailed discussion in text) using the nonlocal 
TPM potential \cite{TPM}. The experimental data were taken from 
Refs.~\cite{exp6,exp7}.} \label{fig3}
\vspace{0.5cm}
\end{figure}

Another way of solving the integro-differential equation (\ref{pareq}) is to divide 
the radial variable into the equal mesh points $\{r_{k}\}$, and to express the
derivative and integral in terms of the wave function $\psi_{lj}(r_k)$ at the mesh points. The integro-differential equation is then transformed to a
set of the linear equations
\begin{equation}
 D_{ik}\psi_{lj}(r_k)=0,\label{lineq}
\end{equation}
where the coefficients $D_{ik}$ are determined from the expansion of the second 
derivative and nonlocal integral over the full set of $\psi_{lj}(r_k)$. 
This method 
has been implemented by Raynal in the DWBA98 code \cite{Raynal} for the elastic and 
inelastic nucleon scattering. 

At variance with these methods, the nonlocality of the nucleon OP 
is treated directly in the calculable $R$-matrix method without involving 
the intermediate steps of an iterative method or using some specific 
transformation of the nonlocal term. Another advantage of the $R$-matrix method 
is the use of the Lagrange-mesh method that allows the direct determination 
of the local (\ref{26}) and nonlocal (\ref{27}) matrix elements. In the present 
work, we have tested the reliability
of the calculable $R$-matrix method by comparing the OM results given by the
$R$-matrix method with those given by the three methods discussed above. 
Fig.~\ref{fig1} presents the OM descriptions of the neutron elastic scattering 
on $^{208}$Pb target at the energies of 14.6, 20, 26, 30.3, and 40 MeV given by the 
four different methods of solving Eq.~(\ref{pareq}), using the same nonlocal 
PB potential \cite{Perey}. One can see that all the considered methods give 
nearly the same differential scattering cross sections over the whole angular 
range, which are indistinguishable in the logarithmic scale. In the case of the 
TPM potential \cite{TPM}, a nonlocal imaginary volume term was added and all the 
parameters have been fitted to give a proper OM description of the elastic
neutron and proton scattering over a wide range of energies. From the OM results 
for the elastic angular distribution of the neutron and proton elastic scattering 
on $^{208}$Pb target shown in Fig.~\ref{fig2} and Fig~\ref{fig3} one can see 
that the calculable $R$-matrix method also reproduces the OM results given by 
the three referred methods.  
\begin{figure}[bht]
\vspace{-1.0cm}
\hspace{-1.0cm}
\includegraphics[scale=0.41]{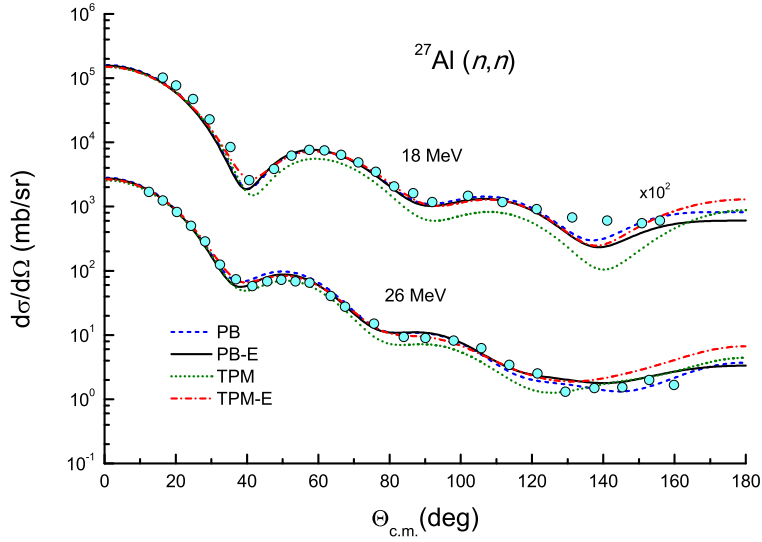} \vspace{-0.4cm}
\caption{OM descriptions of the neutron elastic scattering on $^{27}$Al target 
at the energies of 18 and 26 MeV given by the calculable $R$-matrix method 
using the nonlocal PB and TPM potentials \cite{Perey,TPM} and their 
energy dependent PB-E and TPM-E versions \cite{Lovel}. The experimental data 
were taken from Ref.~\cite{exp8}.} \label{fig4}
\vspace{-0.2cm}
\end{figure}
\begin{figure}[bht]
\vspace{-1.0cm}
\hspace{-1.0cm}
\includegraphics[scale=0.41]{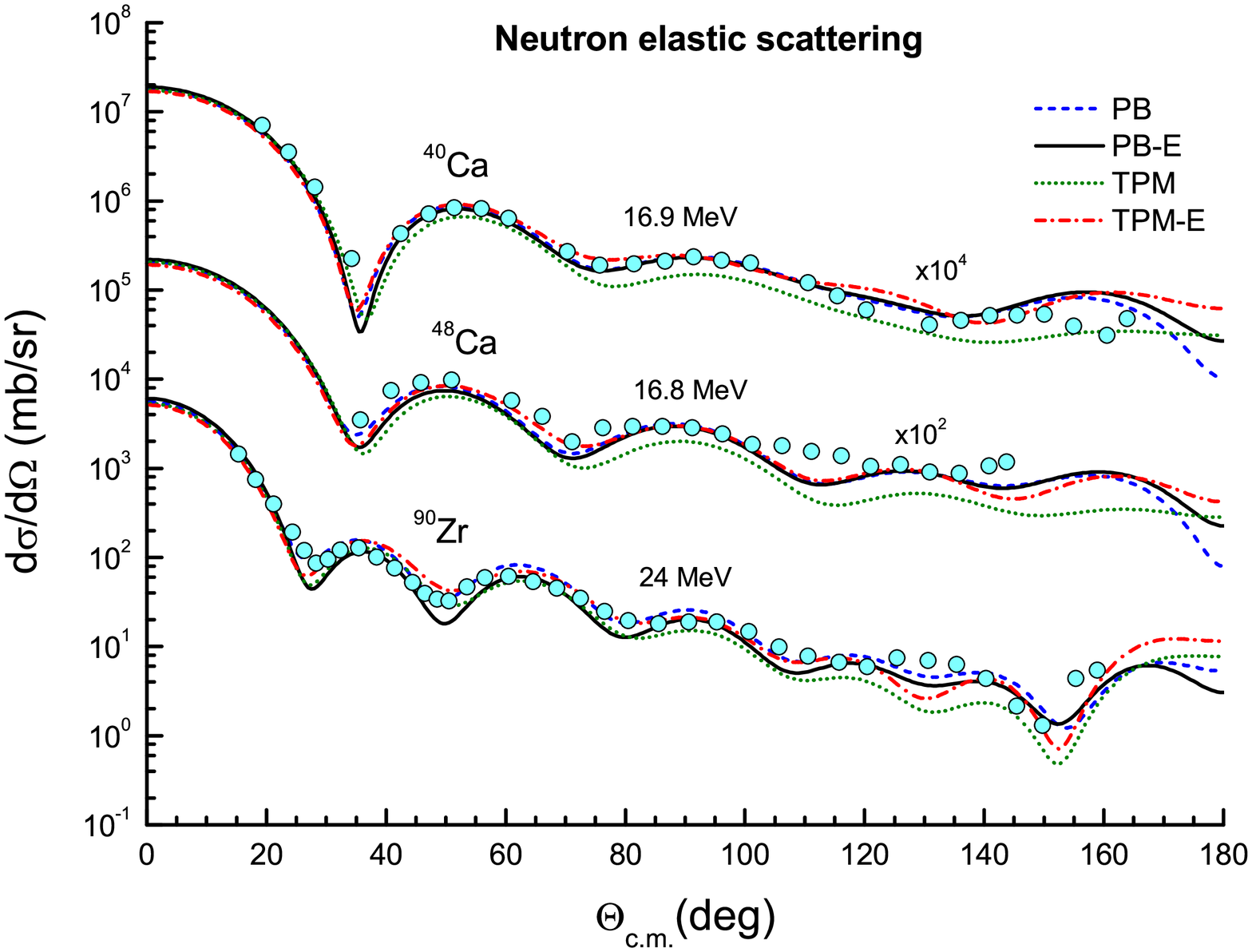} \vspace{-0.3cm}
\caption{OM descriptions of the neutron elastic scattering on $^{40}$Ca,
$^{48}$Ca, and $^{90}$Zr targets at the energies of 16.9, 16.8, and 24 MeV
respectively by the calculable $R$-matrix method using the nonlocal PB and 
TPM potential \cite{Perey} and its energy dependent version \cite{Lovel}. 
The experimental data were taken from Refs.~\cite{exp9,exp10,exp11}.} 
\label{fig5}
\vspace{-0.5cm}
\end{figure}
\begin{figure}[bht]
\vspace{-1.0cm}
\hspace{-1.0cm}
\includegraphics[scale=0.42]{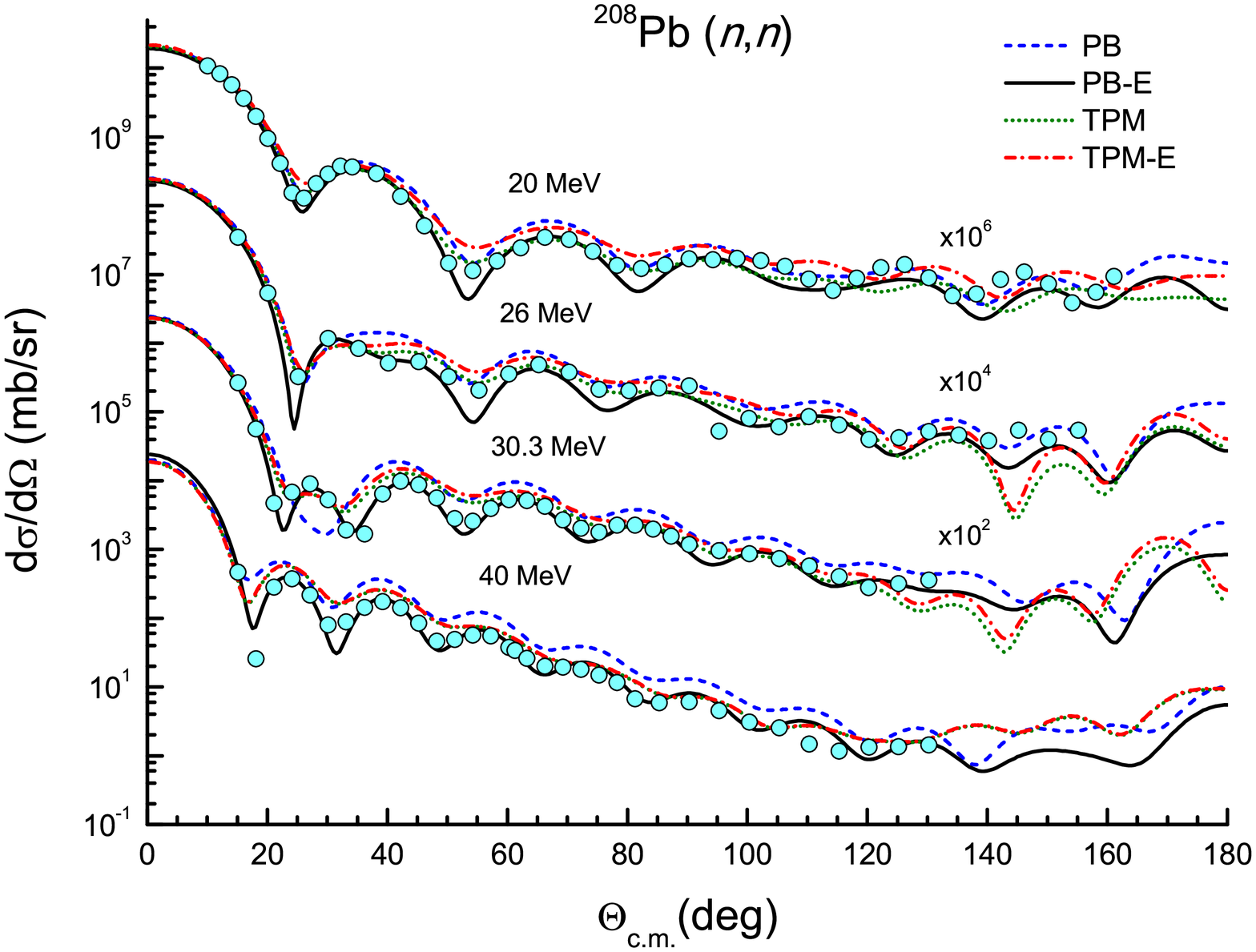} \vspace{-0.3cm}
\caption{OM descriptions of the neutron elastic scattering on $^{208}$Pb 
target at the energies of 20, 26, 30.3, and 40 MeV given by the calculable 
$R$-matrix method using the nonlocal PB and TPM potential \cite{Perey} and 
its energy dependent version \cite{Lovel}. The experimental data were taken 
from Refs.~\cite{exp2,exp3,exp4,exp5}.} \label{fig6}
\end{figure}

Because the parameters of the PB potential were fitted to reproduce the data 
of the neutron elastic scattering on $^{208}$Pb target at energies below 15 MeV, 
it gives a worse OM description of the data at higher energies. That is the reason 
the parameters of the TPM potential were searched \cite{TPM} to obtain a good OM 
description of the elastic nucleon scattering data at the energies of around 10 to 
30 MeV. Our OM results have shown that the TPM potential indeed gives a better OM 
description of the neutron elastic scattering data at 30.3 and 40 MeV (compare the 
lower parts of Fig.~\ref{fig1} and Fig.~\ref{fig2}). From the microscopic
point of view, the imaginary part of the nucleon OP originates mainly from 
the couplings between the elastic scattering channel and open nonelastic 
channels, expressed through the dynamic polarization term in Eq.~(\ref{Feshbach}). 
Because DPP is strongly nonlocal and energy dependent, it is necessary to include 
an explicit energy dependence into the imaginary part of the PB and TPM nonlocal 
potentials as done recently by Lovell {\it et al.} \cite{Lovel}, where the energy 
dependent parameters of the OP were adjusted to fit the neutron elastic scattering 
data on $^{40}$Ca, $^{90}$Zr, and $^{208}$Pb targets at the energies of 5 to 40 
MeV (see PB-E and TPM-E parameters in Table II of Ref.~\cite{Lovel}). The OM 
results obtained with the calculable $R$-matrix method for the neutron elastic 
scattering on $^{27}$Al, $^{40}$Ca, $^{48}$Ca, $^{90}$Zr, and $^{208}$Pb targets 
at the energies of 16.8 to 40 MeV using the original PB, TPM potentials and 
their energy dependent PB-E, TPM-E versions are compared with the 
data in Fig.~\ref{fig4}, Fig.~\ref{fig5}, and Fig.~\ref{fig6}. In all cases, the 
energy dependent PB-E and TPM-E nonlocal potentials give a better OM description 
of the data compared to the original PB and TPM potential, especially, in 
Fig.~\ref{fig4} the good description of the data for $^{27}$Al is obtained at 
both energies although these data sets were not included in the OM search of 
the energy dependent OP parameters in Ref.~\cite{Lovel}. In Fig.~\ref{fig5}, 
the energy dependent nonlocal potentials also give a better OM description of 
the data for $^{40}$Ca, $^{48}$Ca, and $^{90}$Zr compared to the original PB 
and TPM potentials, this behavior is more clearly seen by the TPM-E 
potential compared to TPM potential. The similar result is obtained in
Fig.~\ref{fig6}, where one can see the improved description of the 
data by the energy dependent PB-E nonlocal potential at the energies 
of 30.3 and 40 MeV.
\begin{figure}[bht]
\vspace{-1.0cm}
\hspace{-1.0cm}
\includegraphics[scale=0.42]{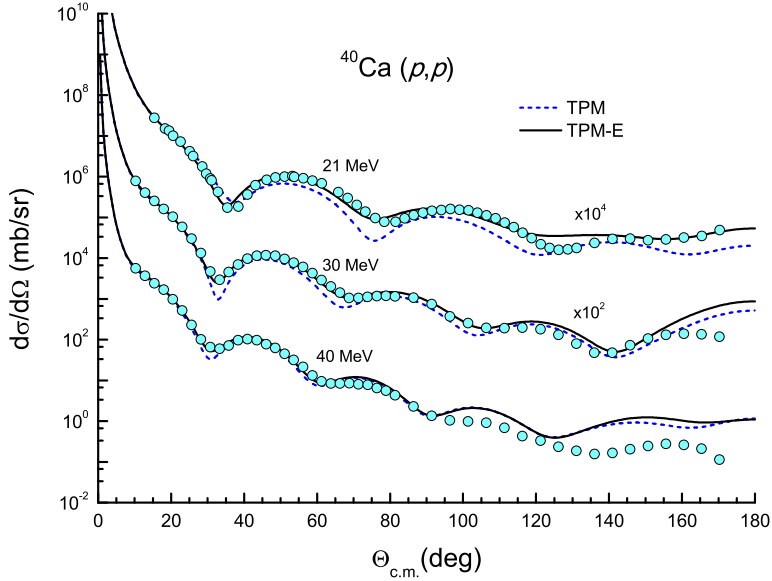} \vspace{-0.3cm}
\caption{OM descriptions of the proton elastic scattering on $^{40}$Ca 
target at the energies of 21, 30, and 40 MeV given by the calculable $R$-matrix
method using the nonlocal TPM potential \cite{TPM} and its
energy dependent TPM-E version \cite{Lovel2}. The experimental data
were taken from Refs.~\cite{exp12}.} \label{fig7}
\end{figure} 

Recently, the energy dependent version TPM-E has been extended to the OM 
study of the elastic proton scattering on $^{40}$Ca, $^{90}$Zr, and $^{208}$Pb
targets at energies of E $\approx$ 10$-$45 MeV \cite{Lovel2}. We have used 
here the TPM nonlocal potentials and the new energy dependent TPM-E version 
in the OM calculation of the proton elastic scattering on $^{40}$Ca target 
at 21, 30, and 40 MeV by the $R$-matrix method. One can see in Fig.~\ref{fig7} 
that the energy dependent TPM-E nonlocal potentials also give a better OM 
description of the data at three energies compared to the original TPM potential. 

\section*{Summary}
The applicability of the calculable $R$-matrix method in the OM calculation 
of the elastic nucleon scattering with a nonlocal nucleon OP
has been explored in the OM analysis of the elastic nucleon scattering on 
$^{27}$Al, $^{40}$Ca, $^{48}$Ca, and $^{208}$Pb targets at the incident 
energies up to 40 MeV, using the phenomenological nonlocal nucleon OP 
proposed by Perey and Buck \cite{Perey}, and the two recent versions of the 
PB parametrization \cite{TPM,Lovel,Lovel2}. 
 
The comparison of the OM results given by the calculable $R$-matrix 
method \cite{desco} with those given by the three other methods 
\cite{NLOM2,Titus2,Raynal} confirms that the calculable $R$-matrix method 
is an efficient, alternative method to treat the nonlocality of the nucleon OP. 
The direct evaluation of the nonlocal term of the OP based on the Lagrange-mesh 
method can be applied further in the OM study using the microscopic nucleon OP 
given by the folding model calculation that treats the nonlocal exchange 
term exactly \cite{Ken01}. 

Based on the present results, it should be possible to use the nonlocal, 
microscopic \nA potential in the $R$-matrix study of the nucleon radiative 
capture at the astrophysical energies. This will be the subject of our upcoming 
research.

\section*{ACKNOWLEDGMENT }
The authors thank Pierre Descouvemont for his helpful discussions and comments
on the calculable $R$-matrix method \cite{desco} and the computer code \cite{desco2}. 
The present research has been supported by Vietnam Atomic Energy Institute (VINATOM) 
under the grant number CS/18/04-01.

\end{document}